\documentclass[conference, letterpaper]{IEEEtran}

\usepackage{amsmath}

\usepackage{array}
\usepackage{ulem}

\usepackage{url, xcolor}

\usepackage{multirow}
\usepackage{amsfonts}
\usepackage{fancybox, nccmath}

\usepackage{lipsum}  
\usepackage[inline]{enumitem}
\usepackage{tablefootnote}
\usepackage{hyperref}
\usepackage{wrapfig}
\usepackage{cleveref}
\usepackage{float} 
\usepackage{graphicx}
\usepackage{textcomp}

\newcommand{\red}[1]{\textcolor{black}{#1}}

\begin{document}

\title{Can we trust our energy measurements? \\ \LARGE{A study on the Odroid-XU4.}}

\author{\IEEEauthorblockN{Julius Roeder}
\IEEEauthorblockA{University of Amsterdam \\
Amsterdam, The Netherlands \\
Email: j.roeder@uva.nl}
\and
\IEEEauthorblockN{Sebastian Altmeyer}
\IEEEauthorblockA{University of Augsburg \\
Augsburg, Germany \\
Email: altmeyer@es-augsburg.de}
\and
\IEEEauthorblockN{Clemens Grelck}
\IEEEauthorblockA{University of Amsterdam \\
Amsterdam, The Netherlands \\
Email: c.grelck@uva.nl}
}


\maketitle

\begin{abstract}
  IoT devices, edge devices and embedded devices, in general, are ubiquitous. The energy consumption of such devices is important both due to the total number of devices deployed and because such devices are often battery-powered. Hence, improving the energy efficiency of such high-performance embedded systems is crucial. The first step to decreasing energy consumption is to accurately measure it, as we base our conclusions and decisions on the measurements. 
  Given the importance of the measurements, it surprised us that most publications dedicate little space and effort to the description of their experimental setup.

  One variable of importance of the measurement system is the sampling frequency, e.g. how often the continuous signal's voltage and current are measured per second. In this paper, we systematically explore the impact of the sampling frequency on the accuracy of the measurement system. We measure the energy consumption of a Hardkernel Odroid-XU4 board executing nine Rodinia benchmarks with a wide range of runtimes and options at 4kHz, which is the standard sampling frequency of our measurement system. We show that one needs to measure at least at 350Hz to achieve equivalent results in comparison to the original power traces. Sampling at 1Hz (e.g. Hardkernel SmartPower2) results in a maximum error of 80\%. 
  
\end{abstract}

\section{Introduction}

Energy consumption is one of the most important design criteria for battery-powered systems. Thus, it is not surprising that decreasing energy consumption from the software side is an important topic in various research fields such as IoT, edge computing, cyber-physical systems and embedded systems (e.g. \cite{cloutier2016raspberry, roeder2021energy, guo2019energy, liu2015energy, pricopi2013power, balsini2019modeling, zeng2009practical, odyurt2021power, isuwa2019teem, brouwers2014neat}; for a survey see \cite{sheikh2018energy}).

A crucial part of energy related research is measuring the energy consumption in order to show tangible improvements on real hardware. To measure the energy consumption of a device we need to measure voltage and current, two continuous signals. Continuous signals are measured in discrete intervals at a given sampling rate. From a theoretical point of view we need to measure at twice the highest frequency desired to be measured (Nyquist rate \cite{oppenheim2001discrete}) otherwise the time series might be distorted. 
However, from a practical point of view it is unclear what the highest desired frequency in this case is.

In general we find that authors and reviewers place little importance on the measurement setup, as papers do not report the setup or lack details on the devices and methods used, e.g. \cite{guo2019energy, liu2015energy, pricopi2013power, balsini2019modeling, zeng2009practical,imes2015poet}. Publications that do report the measurement system used, do not investigate or consider the impact of the measurement setup on the accuracy of the measurements. 
For example, \cite{isuwa2019teem} naturally used the energy measurement system (SmartPower2\footnote{\url{https://www.hardkernel.com/?s=smartpower2&post_type=product&lang=en}}) provided by the manufacturer of their target board (Odroid-XU4). According to the publication the SmartPower2 measures at 1Hz. Additionally, we could not find any studies on the measurement error of the SmartPower2. 
In this paper we raise strong doubts about the reliability of low frequency measurements. As a community, that makes decisions based on energy consumption, we must know that our experimental setups are reliable. 

To the best of our knowledge no prior paper has investigated the correlation between sampling frequency and accuracy of energy measurement systems for high-performance embedded systems. Thus, in this paper we systematically investigate the impact of the sampling frequency on the energy measurement accuracy. 
More specifically we measure the energy consumption of an Odroid-XU4 executing a variety of benchmarks. The measurement system used samples at a high rate; the original power-traces can then be downsampled. We then compare the downsampled traces against the original traces. That way we can alter the sampling frequency of the voltage and current measurements, while keeping all other variables equal.

The paper is organised as follows. In \Cref{sec:background_methodology} we provide background information and detail our methodology. \Cref{sec:results} covers our results and discussion. Then in \Cref{sec:related_work} we discuss related work. Finally, we present our conclusion in \Cref{sec:conclusion}.

\section{Background \& Methodology} \label{sec:background_methodology}
In order to investigate the importance of sampling frequency we need a measurement system, a target system, programs to measure and a way to compare different sampling frequencies. In this section we start with a short discussion about power measurements in general. We then dive into the importance of sampling frequency. Next we introduce our experimental setup and the benchmarks used. Lastly, we detail the  statistical tests needed.

\subsection{Power Measurements}
Power measurements can be done at the AC source or at the DC source. Discussing and comparing the advantages of either method is beyond the scope of this paper. However, in general the AC-DC converter (i.e. power supply) will have some inefficiencies, and measuring after the converter (i.e. at DC) disregards the loss. Furthermore, the loss can fluctuate with the load, i.e. power supplies are most efficient at a given load and have lower efficiencies at lower/higher loads. An additional reason to measure after the converter is that the energy consumption is most crucial for battery powered systems, which use DC. 

\red{For an overview of different DC measurement methods see \cite{nakutis2009embedded} and \cite{hergenroder2012energy}. In this paper we consider the shunt resistor method which observes the voltage drop across a resistor as it is widely used.} We place the resistor in series with the load. And as we know the resistor value, Ohm's law can be applied to calculate the current of the load. Furthermore, we can place the resistor before the load (high side) or after the load (low side). Low side sensing is cheaper as the amplifier is simpler but has some disadvantages in comparison to high side. More specifically low side sensing is sensitive to ground disturbances and (in this case less importantly) cannot detect fault conditions. Hence, it is mostly used in mass production systems \cite{low_vs_high_side}. Therefore, we will focus on high side sensing.

\red{The resistive current sensing method can be deployed directly on a target board, i.e. the board comes with an integrated power measurement function (e.g. Odroid-XU+E\footnote{\url{https://www.hardkernel.com/shop/odroid-xue/}} used in \cite{imes2015poet}). Or the method can be deployed on a separate device such as the SmartPower2 or Qoitech Otii\footnote{\url{https://www.qoitech.com/otii/}}. Onboard sensors are polled from the target system itself and can be polled at different frequencies. Additionally, onboard sensors are intrusive as the polling of the sensors impacts the energy consumption of the target.}

The voltage drop across the resistor is amplified and then converted using an Analogue-Digital-Converter (ADC). Current sense amplifiers such as the TI INA250\footnote{\url{https://www.ti.com/product/INA250}} can be used in combination with an ADC. The ADC then digitises the information for further analysis. 

\red{Once we obtained the voltage and current readings we can calculate the power (Watt). Multiple power readings result in a power trace and as we know the time between different power readings we can calculate the area under the trace, resulting in the energy consumption (Joule).}

\subsection{Sampling Frequency}
Continuous signals cannot be converted to digital information continuously, instead we have to measure them at discrete intervals. The accuracy of the measurements heavily depends on the sampling frequency. 
In theory to reproduce an (AC) power signal one needs to measure voltage and current at four times of the highest sinusoidal frequency \cite{Turgel1975}.
However, the DC consumption is not sinusoidal and instead alternates with requirements of the load. In the case of a micro-controller the current requirements change with for example the Dynamic Voltage and Frequency Scaling (DVFS) settings, instructions per clock and the actual instructions being executed \cite{vasilakis2017modeling}. Thus, the required sampling frequency depends on the length of the program being executed and the instruction mix. 

\subsection{Setup and Target system}
High-performance embedded systems like the Odroid-XU4 and the NVidia Jetson Nano are all relatively similar with respect to the clock frequency and CPU architecture. In this paper we use the Odroid-XU4 board \cite{odroid123} as an example target system. It is an octa-core system with 4 big cores (Cortex-A15), 4 LITTLE cores (Cortex-A7) and a Mali-GPU (T628 MP6). 
The two separate core clusters and the GPU all form individual voltage islands (i.e. 3 voltage islands). The voltage and the frequency can be set separately for each voltage island. The Odroid-XU4 runs an RT-patched Linux.

The Odroid-XU4 is accompanied by an energy measurement system called the SmartPower2. However, due to the low sampling frequency (1Hz) we decided to not use the system. Instead, we measure the energy consumption of the Odroid-XU4 with the Qoitech Otii on the high side. The Otii has a maximum measurement error of 0.1\% + 150µA (i.e.  at higher currents the error is approaching 0.1\%) and has a sampling frequency up to 4kHz. \red{The main criticism of the shunt resistor method is that a single shunt is only useful in a limited current range \cite{buschhoff2014mimosa, hergenroder2012energy}. The Otii has multiple shunts to measure very low currents (10~{\textmu}A with 0.6\% error) up to 5A peaks. It measures across all shunts at the same time, thus switching current range (i.e. between shunts) does not result in any lost data points.}

\begin{figure}
    \centering
    \includegraphics[width=\linewidth]{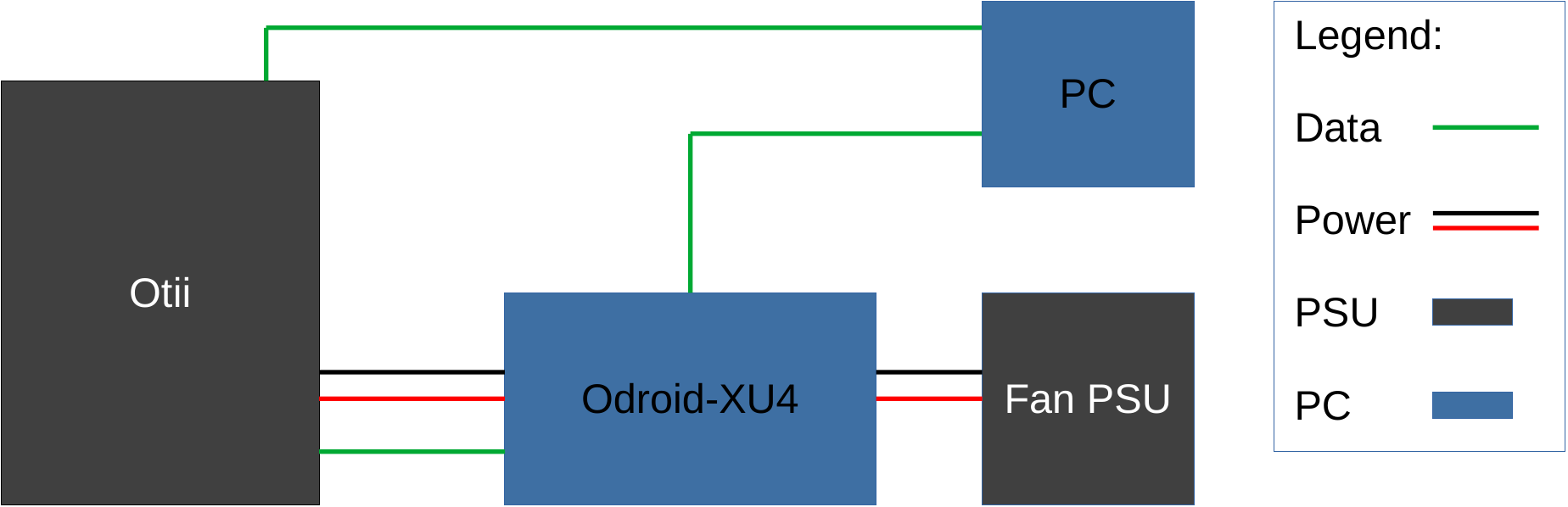}
    \caption{Measurement setup including: Qoitech Otii, Odroid-XU4, Fan power supply, and PC}
    \label{fig:measurement_setup}
\end{figure}

\Cref{fig:measurement_setup} shows our setup. The Odroid-XU4 receives its power from the Otii and is at the same time connected via the UART pins to the Otii. This means that the power measurements can be directly linked to messages sent by the Odroid-XU4. The fan of the Odroid-XU4 is powered via a separate circuit, and thus does not affect the power measurements of the Odroid-XU4. 
Before each set of measurements, we calibrate the Otii. Additionally, we warm up all connected components by executing the \textit{heartwall} benchmark 50 times.

\subsection{Downsampling}
The Otii samples at 4kHz. Instead of either forcing a lower sampling rate or using a device with a lower sampling rate we downsample the results.  That means if we sample at 4kHz but want a sampling rate of 2kHz we only take into account every second measurement. Thus, sampling unrelated factors do not play a role (e.g. different measurement error on a different measurement device). In this paper we investigate 22 sampling rates (in Hz: 1, 2, 3, 4, 5, 10, 20, 30, 40, 50, 100, 150, 200, 250, 300, 350, 400, 500, 600, 800, 1000, 2000, 4000).

\subsection{Benchmarks}
We use the Rodinia benchmark suite \cite{che2009rodinia} as target programs/tasks. The suite offers a range of targets (C, OpenCL, Cuda), different algorithms \& workloads and is widely used. The suite's benchmark selection was inspired by Berkeley's dwarf taxonomy \cite{asanovic2006landscape}. Each benchmark can further be adjusted via its input parameters. This leads to a large range of run-times and processor loads. 

We use nine benchmarks (backpropagation, BFS, Heartwall, Hotspot, Kmeans, LU-Decomposition, Nearest Neighbour, NW, SRAD) out of the suite as they can be executed on the Odroid-XU4 with minimal adaptations. The other benchmarks would have required significant changes to the code. Besides the input parameters we also vary the target DVFS settings and the target core. We measure all benchmarks on the LITTLE cores, on the big cores and on the GPU (i.e. OpenCL version). However, there are two benchmarks (BFS and SRAD) which were only measured on the big and on the LITTLE cores because the OpenCL versions did not work on the Odroid-XU4. This leads to a total of 842 unique benchmark/target/DVFS combinations. For each combination we collected 50 power traces, thus, in total we collected 42100 power traces.

\red{The resulting dataset is available for download \footnote{\url{https://doi.org/10.21942/uva.19665564.v1}}\cite{Roeder2022}. Additionally, the repository containing the analysis scripts is also available \footnote{\url{https://bitbucket.org/uva-sne/energymeasurementanalysis/}}.}

\subsection{Statistical equivalence testing}

We measure a non-deterministic system (out-of-order pipeline etc.). Additionally, the measurement system is not perfect and contains some noise. Thus, we repeat measurements for each combination, as there is not a single ''correct`` value. That also means that downsampling a single time-series and then calculating the error will give an indication of how much worse a lower frequency is. However, this approach does not offer a statistical indication. Therefore, we need to analyse all sets and their downsampled counterparts with statistical tests. 

In a regular two-sided t-test we test if two samples are different. The null hypothesis is that there is no difference ($\mu_D$) between two samples (\Cref{eq:null_hyp_ttest}).

\begin{align}
    H0: \mu_D = 0 \label{eq:null_hyp_ttest}\\
    H1: \mu_D != 0 \label{eq:alternative_hyp_ttest}
\end{align}

If the t-test indicates a significant difference (e.g. p-value smaller than 0.05) then we can reject the null hypothesis and accept the alternative hypothesis that the two samples are different (\Cref{eq:alternative_hyp_ttest}). Thus, a t-test offers evidence in favour of the alternative hypothesis at a given confidence level (e.g. 99\%). If the t-test is not significant, this is often counted as support for the null hypothesis, i.e. that there is no difference between the samples or that there is no effect. However, often a non-significant test result is the result of limited statistical power. Thus, it is impossible to know whether a non-significant result indicates equivalence (absence of an effect) or only false equivalence and is lacking statistical power \cite{quertemont2011statistically}. 

Instead of proving the absence of an effect, we can show that the likelihood of an effect being smaller than a given (low) value to be significant, this is called equivalence testing. To test for equivalence between two samples we use a method called \textit{Two One Sided T-tests} (TOST) \cite{quertemont2011statistically}. As a TOST consists of two tests, it has two null hypotheses (\Cref{eq:null_hyp_TOST1}) and (\Cref{eq:null_hyp_TOST2}). The first test is used to determine if the difference between the two samples ($\mu_D$) is smaller than the accepted lower bound ($-M$). The second one tests if the difference is larger than the upper bound $M$. 

\begin{align}
    H0_1: \mu_D < -M \label{eq:null_hyp_TOST1}\\
    H0_2: \mu_D > M \label{eq:null_hyp_TOST2}
\end{align}

\begin{figure}
    \centering
    \includegraphics[width=\linewidth]{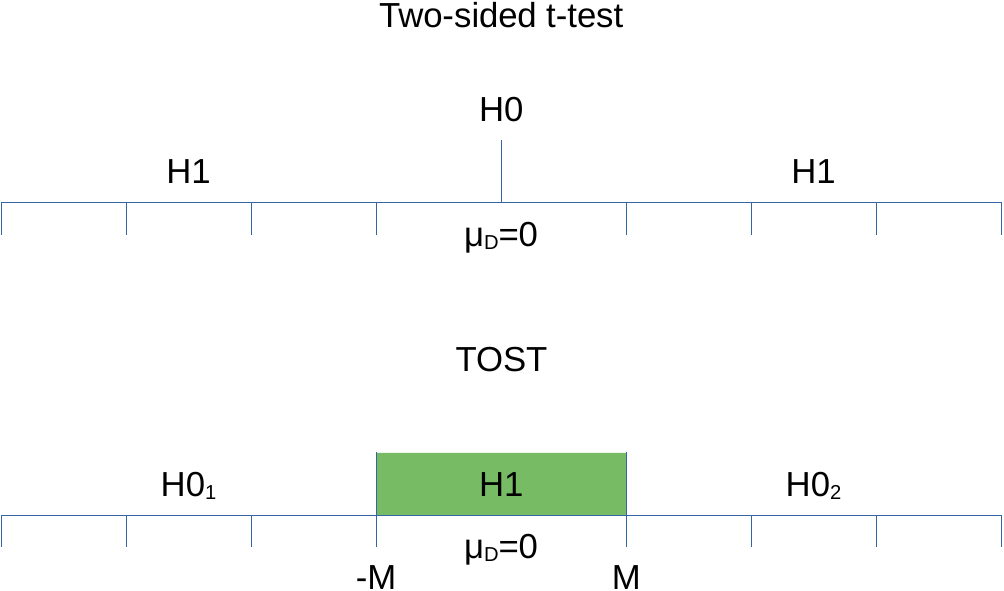}
    \caption{Comparison of a two-sided t-test and a TOST.}
    \label{fig:equivalence_vs_ttest}
\end{figure}

Combining both test results in the alternative hypothesis (\Cref{eq:alternative_hyp_TOST}) that $\mu_D$ falls between $-M$ and $M$. Thus, if both t-tests are rejected, we have support for the alternative hypothesis that the difference between the two samples is smaller than a chosen $M$ \cite{ncss}. \Cref{fig:equivalence_vs_ttest} visualises the difference between a normal t-test and a TOST.

\begin{align}
    H1: -M < \mu_D < M \label{eq:alternative_hyp_TOST}
\end{align}

The majority of our 842 measurement sets are not normally distributed (76.0\%) according to both the Shapiro-Wilk test \cite{shapiro1965analysis} and D’Agostino-Pearson’s test \cite{d1973tests}. 
Therefore, we use a non-parametric TOST based on Wilcoxon's Signed Rank test \cite{wilcoxon1992individual}. We do all tests at a 99.9\% confidence ($\alpha=0.1\%$).

One major difference between a standard t-test and an equivalence test is that one needs to determine what (low) difference ($M$) is acceptable (i.e. considered to be less than a noteworthy effect). We analyse the impact of 8 ''acceptable error`` levels (20\%, 10\%, 8\%, 6\%, 4\%, 2\%, 1\%, 0.5\%) and what sampling level is required to achieve equivalence at that level across all 842 experiment combinations.

\section{Results \& Discussion} \label{sec:results}
The 42100 power trace time-series can be analysed in multiple different ways. \Cref{tab:general_stats} summarises basic statistics of all power traces and shows that our benchmarks/target/DVFS combinations cover a wide range of run times and power. Overall we observe that the downsampled traces mostly resulted in a power consumption underestimation (98.9\% of the cases) and in very few cases of overestimation (1.1\%). 

\begin{table}[htb]
\centering
\caption{Summary statistics for all benchmark executions.}
\label{tab:general_stats}
\begin{tabular}{l|ll}
     & Runtime (s) & Power (W) \\ \hline
Mean & 9.87        &    2.99       \\
Min  & 0.90        &    1.82       \\
Max  & 48.15       &    8.44      
\end{tabular}
\end{table}

\begin{figure}[htb]
    \centering
    \includegraphics[width=8cm]{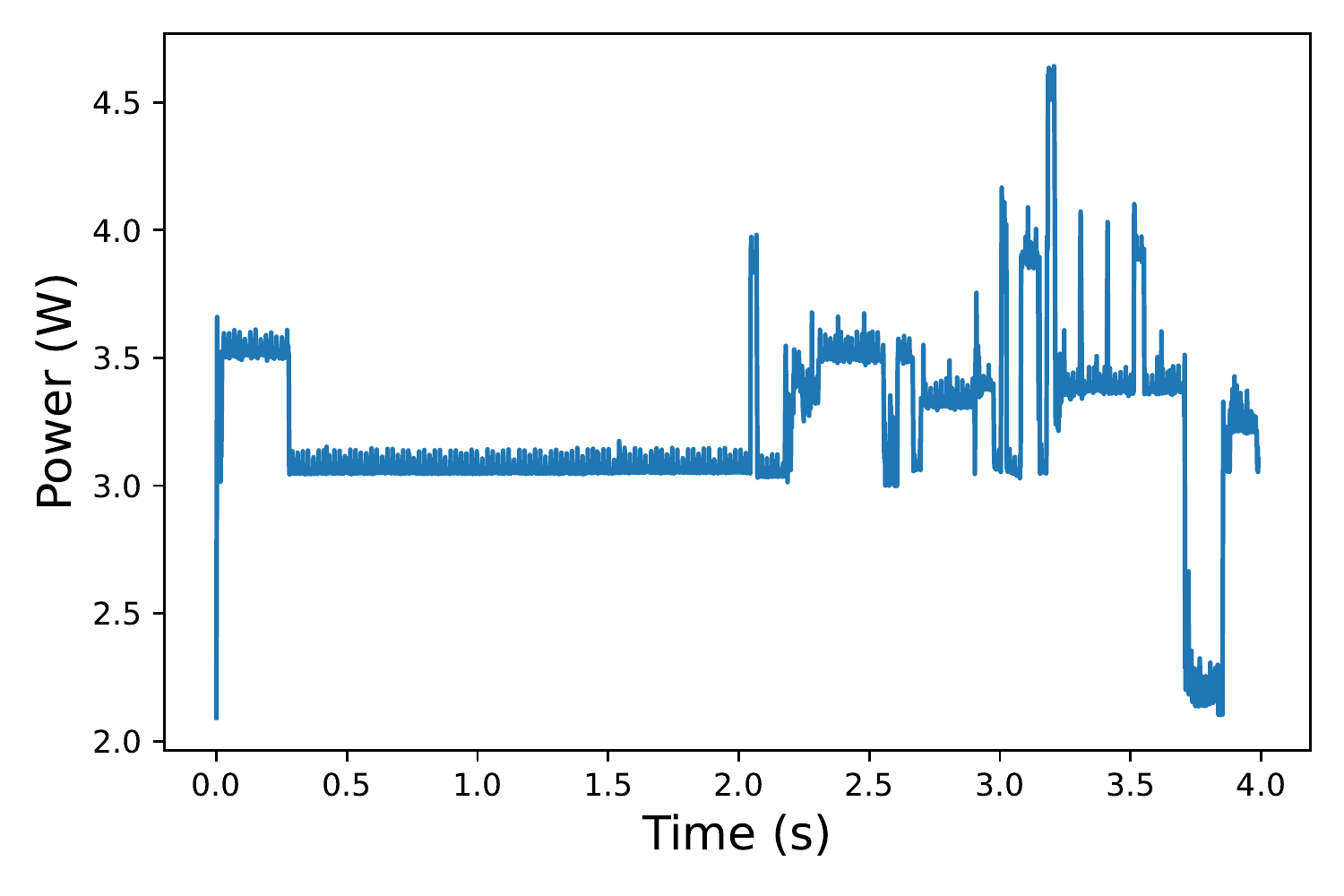}
    \caption{Original power trace sampled at 4000Hz.}
    \label{fig:power_trace}
\end{figure}

\begin{figure}[htb]
    \centering
    \includegraphics[width=8cm]{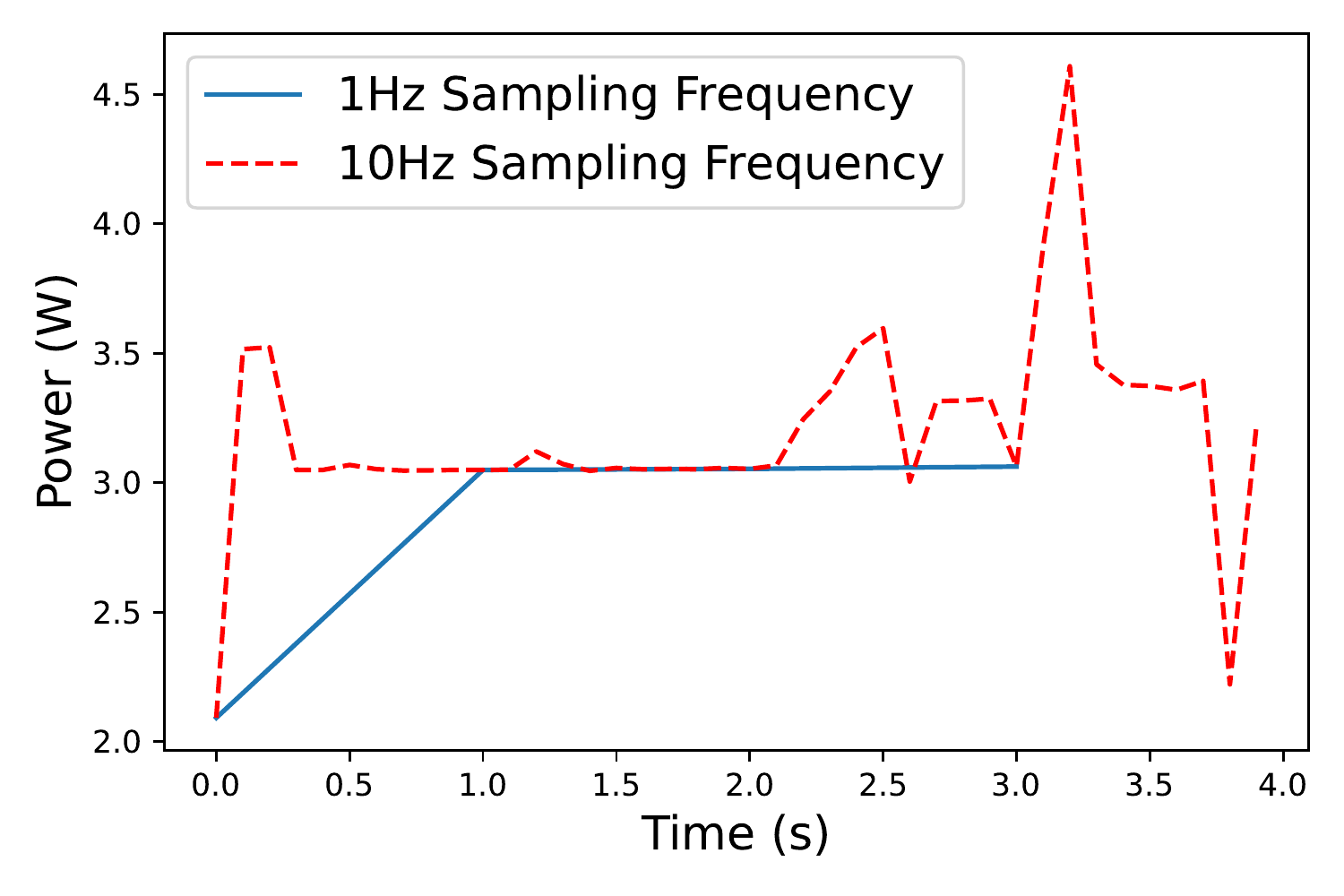}
    \caption{Downsampled power traces.}
    \label{fig:power_trace_downsampled}
\end{figure}

\Cref{fig:power_trace,fig:power_trace_downsampled} show one of the power traces.  \Cref{fig:power_trace} shows the original power trace at the full sampling frequency of 4kHz and \Cref{fig:power_trace_downsampled} shows two downsampled versions. The solid blue line in \Cref{fig:power_trace_downsampled} shows how the power trace looks like if we had sampled at 1Hz. In comparison to the original trace we can see that it misses a majority of the data. Furthermore, it also misses data on the last second completely, as the execution time was 3.98 seconds. \red{It is possible to make up for the last missed measurement by either taking the measurement at second 4 or by using the last known measurement. Either method will still lead to a significant error.} The dashed red line shows the same power trace but downsampled to 10Hz. It already has a lot more detail than the 1HZ line but still misses a significant part of the signal. 

\begin{figure}[htb]
    \centering
    \includegraphics[width=8cm]{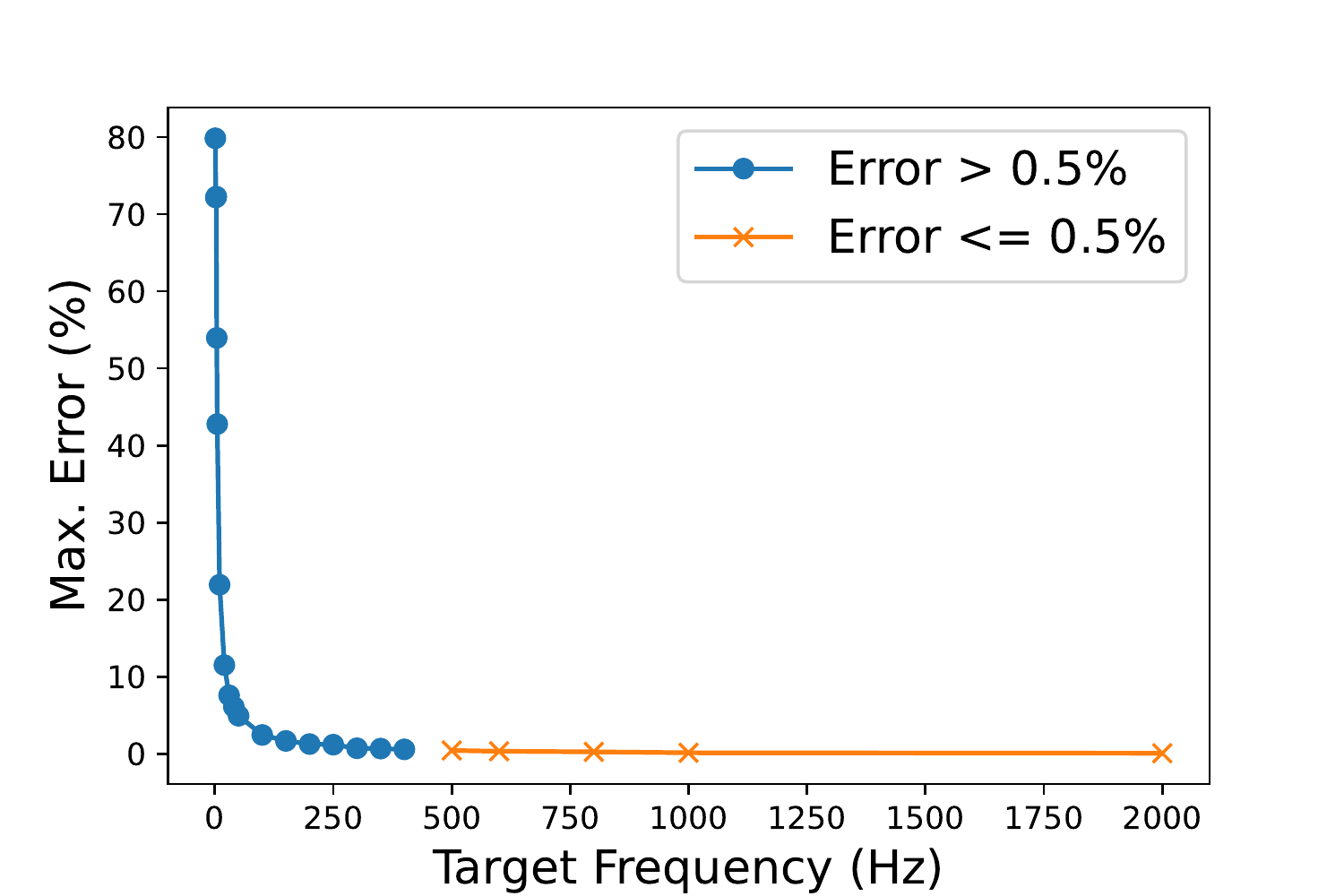}
    \caption{Maximum error rate at each artificial frequency across all 842 experiment sets.}
    \label{fig:max_error_per_frequency}
\end{figure}

\Cref{fig:max_error_per_frequency} shows the maximum percentage error between the original \red{energy} measurement and the downsampled measurement for each frequency. Thus, the maximum error observed across all 842 combinations at 1Hz is 80\%. The maximum error only drops below 0.5\% at a sampling frequency of 500Hz.  

\begin{figure}[htb]
    \centering
    \includegraphics[width=8cm]{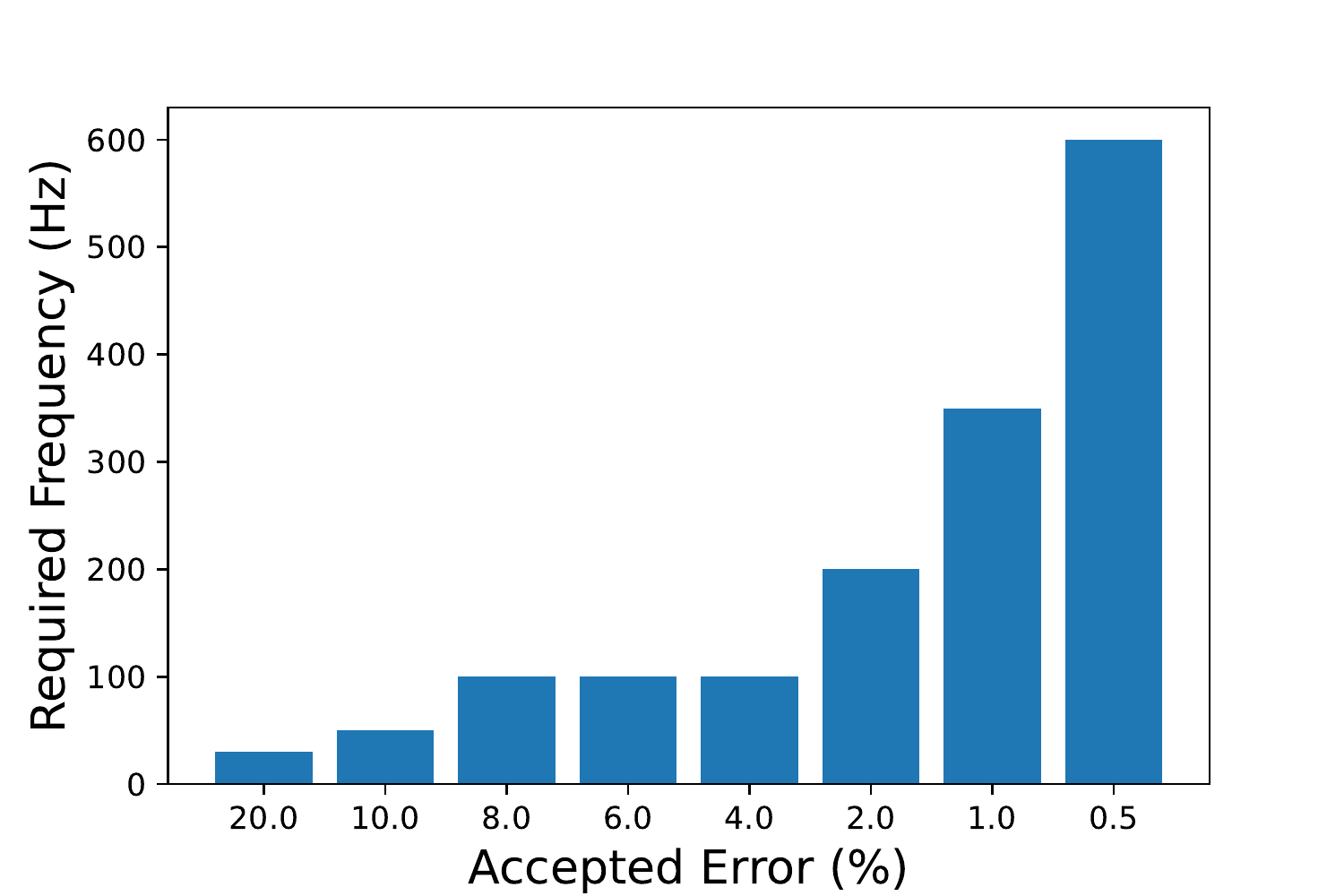}
    \caption{Frequency required to reach equivalence given an acceptable error.}
    \label{fig:equivalence_testing}
\end{figure}

The maximum error only represents a single measurement and does not carry any statistical meaning, which is the reason to employ equivalence testing. \Cref{fig:equivalence_testing} shows the minimum frequency required to achieve equivalent results for all 842 combinations in comparison to the full sampling frequency. Thus, if a measurement error of up to 20\% is acceptable then a 30Hz sampling rate would lead to an equivalent result for all experimental combinations. At an acceptable error of 0.5\%, 600Hz results in an equivalent result. Thus, at a similar level as indicated in \Cref{fig:max_error_per_frequency}. 

\begin{figure}[htb]
    \centering
    \includegraphics[width=8cm]{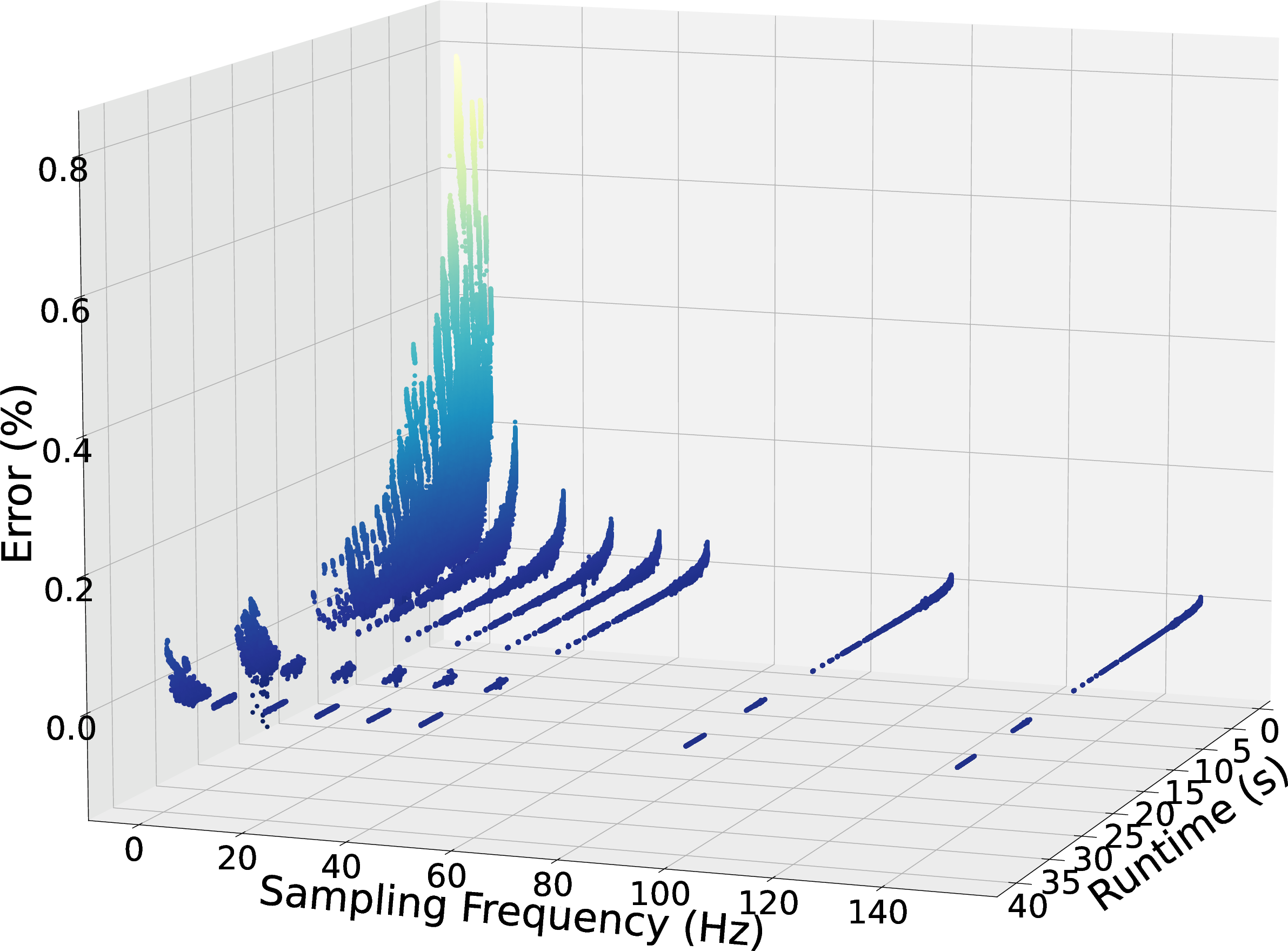}
    \caption{Relation between the error, benchmark run-time and the sampling frequency for all power traces downsampled to between 1Hz and 140Hz.}
    \label{fig:messy_3d}
\end{figure}

\begin{figure}[htb]
    \centering
    \includegraphics[width=8cm]{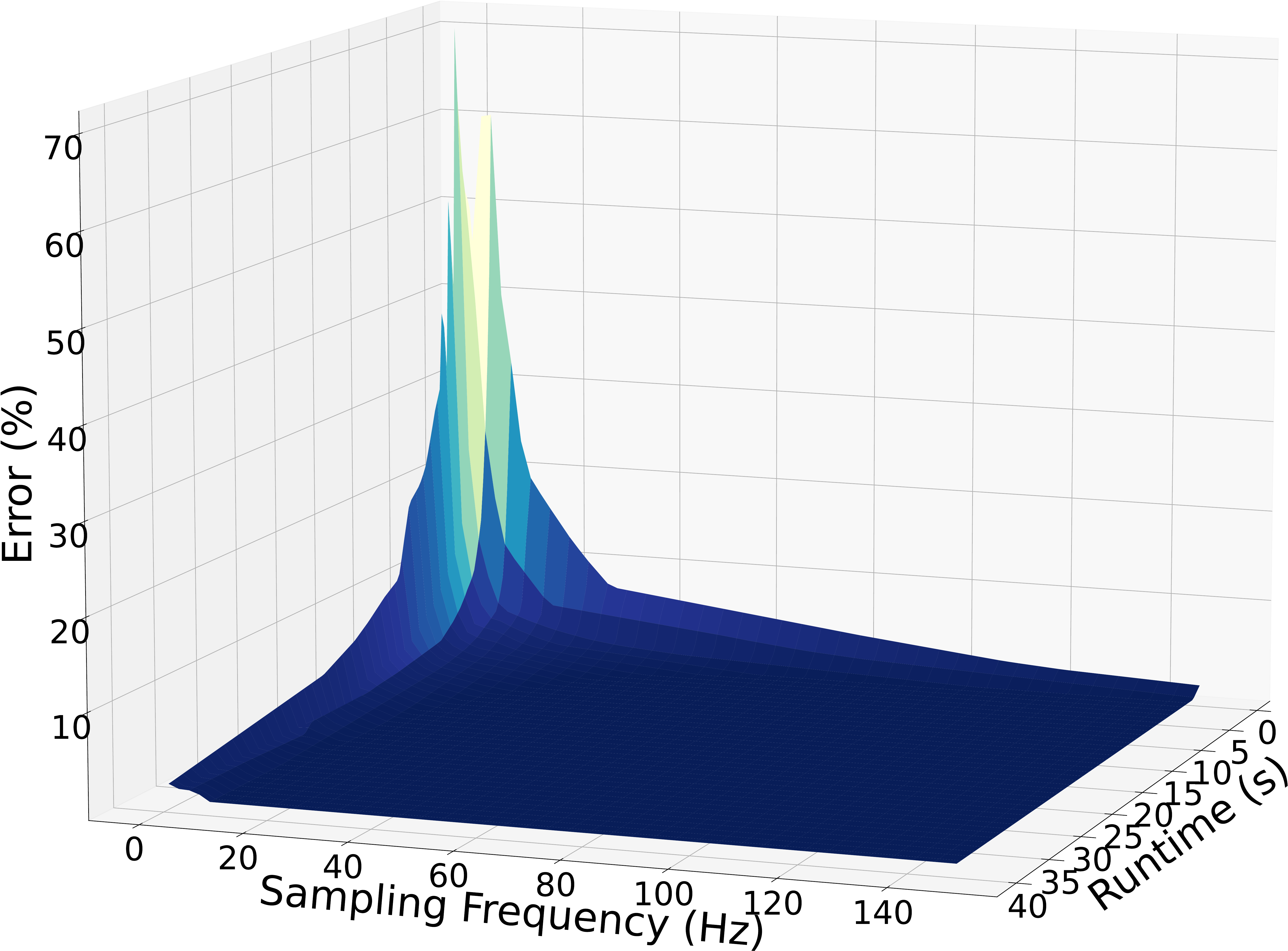}
    \caption{Regression analysis of the error with respect to the benchmark run-time and the sampling frequency.}
    \label{fig:MLPRegressor}
\end{figure}

Lastly, in \Cref{fig:messy_3d,fig:MLPRegressor} we investigate the relation between the error, benchmark run-time and the sampling frequency. Interpreting the 3D graph showing the relation between all three is not straightforward as the resulting graph contains a lot of non-continuous data points (\Cref{fig:messy_3d}). To ease the interpretation, we smooth the data and the relation between the three variables using a polynomial, multi-variable regression based on a Multi-Layer-Perceptron (Scikit-learn: default parameters, hidden layer size = (64, 128, 256, 512)). This also allows us to interpolate the error to other sampling frequencies and run-times. We use 80\% of the data for training. The mean absolute error on the test set is 0.0065. 

We use the regression to predict the error of a measurement given a sampling frequency and run-time. Plotting the regression for the sampling frequency range 1Hz to 140Hz and run-times between 0.5 and 40 seconds results in \Cref{fig:MLPRegressor}. \red{The figure clearly shows that low sampling frequencies lead to poor results for the selected benchmarks even for longer run-times. That means that the selected long running benchmarks contained a significant amount of faster peaks that were missed at a low sampling rate. }
The error for short tasks remains higher even with higher sampling frequencies. As such the results obtained with a SmartPower2 are of limited use in an academic setting.

For this set of benchmarks, input parameters, target platform and DVFS settings a sampling frequency between 350Hz and 600Hz is sufficient (given an error of 1\% and below). However, much shorter programs might need significantly higher sampling rates or one will have to measure the target task in a different way. For example, measuring a very short task (a few CPU cycles) will be missed even at a sampling frequency of 4kHz, thus, artificially inflating the task could work (e.g. a loop).

\section{Related Work} \label{sec:related_work}

Cloutier et. al demonstrate that decreasing the sampling frequency from 100Hz to 1Hz results in significant loss of the power trace detail \cite{cloutier2016raspberry}. However, they do not further investigate the impact of this decrease on the energy measurement accuracy. Additionally, we can show that the accuracy of measurements at 100Hz is significantly lower than at 4kHz. 

Diouri et. al investigate different energy measurement systems for servers \cite{diouri2013solving}. They conclude that higher sampling rates are not necessarily good as they can introduce noise that could mask other trends. However, only because a signal is more noisy doesn't mean that the noise is erroneous and can thus be disregarded for energy measurements. One can always downsample a trace or smooth it to investigate possible hidden trends. Furthermore, server measurements could already be more noisy than high performance embedded systems due to architectural reasons, different target applications and short background tasks. Looking at \Cref{fig:power_trace} we cannot confirm that a high sampling rate masks the trends of an application. Lastly, Diouri et. al do not investigate if the downsampled traces lead to equivalent energy measurements. 

Djupdal et. al \cite{djupdal2021lynsyn} develop a high-performance embedded system oriented energy measurement systems. And in \cite{ilsche2015power} the authors describe two high-sampling frequency power measurement methods (up to 500kHz) for servers and for server components. However, they do not analyse the importance of the sampling frequency and if lower sampling frequencies can achieve similar results. 

\red{Buschoff et. al \cite{buschhoff2014mimosa} and Jiang et. al \cite{jiang2007micro} developed measurement techniques for low-powered embedded systems. They target devices with long sleep times that only consume energy in a few fast bursts. In contrast we focus on high-performance embedded systems that carry out computationally demanding tasks.}

\red{Nakutis et. al \cite{nakutis2009embedded} and Hergenr{\"o}der et. al \cite{hergenroder2012energy} summarise the different power measurement methods and highlight the importance of the sampling frequency. However, neither paper empirically shows the resulting error.}

\section{Conclusion} \label{sec:conclusion}

Research into reducing energy consumption of embedded systems is popular. Hence, we need to measure the energy consumption of embedded systems. However, researchers and reviewers alike often pay little attention and consideration to how to measure energy consumption. One crucial aspect of energy measurements for high-performance embedded systems is the sampling frequency of the analogue signal. 

In this paper we show that for a wide range of Rodinia benchmarks executed on the Odroid-XU4 the minimum sampling rate is 350Hz if a 1\% measurement error is acceptable. Measuring at 1Hz results in errors as high as 80\%. Thus, showing that systems such as the Hardkernel SmartPower2 (measurement system accompanying the Odroid-XU4) cannot be used to draw conclusions and that measurement methods with low sampling rates are only of limited use in an academic setting. Some papers in the area of reducing energy consumption of high-performance embedded systems should be re-evaluated.  

If we want to reliably research and investigate methods for reducing energy consumption we must measure energy consumption accurately. That means that we need to pay more attention to our experimental setup and report our setup accurately. Careless experimental setups lead to two problems: First, we potentially focus too much on the wrong methods (false positive conclusion). Second, we discard methods that do not look promising but are in reality a good option (false negative conclusion). 

In the future we would like to establish theoretical minimum requirements for sampling rate. And work on a community based set of guidelines for energy measurements in the high-performance embedded systems area to avoid such problems and confusion henceforth. 


\section*{Acknowledgements}
We would like to thank the reviewers for their time and feedback.
This work is supported and partly funded by the HiPEAC project which has received funding by the European Union Horizon-2020 research and innovation programme under grant agreement No. 871174 (HiPEAC6 Network). Additionally, this work is partially supported by the European Union Horizon-2020 research and innovation programmes TeamPlay (grant agreement No. 779882) and ADMORPH (grant agreement No. 871259). Lastly, this work is partially supported by CERCIRAS COST Action CA19135 funded by COST Association.

\bibliographystyle{IEEEtran}
\bibliography{bibliography}

\end{document}